\documentclass[preprint,superscriptaddress]{revtex4}

\usepackage{graphicx}
\usepackage{amsmath,amssymb,amsfonts,epsfig}
\usepackage{graphicx}
\usepackage{subfigure,enumerate}
\usepackage{chapterbib}
\usepackage{setspace}

\newcommand{\ups}{V}

\newcommand{\dslash}{\not{\hbox{\kern-2pt $\partial$}}}
\newcommand{\bea}{\begin{eqnarray}}
\newcommand{\eea}{\end{eqnarray}}
\newcommand{\rw}{\rightarrow}

\begin{document}


\title{Beyond power laws: Universality in the average avalanche shape}
\author{Stefanos Papanikolaou}
\affiliation{LASSP, Department of Physics, Clark Hall, Cornell University, Ithaca, NY 14853-2501}
\author{Felipe Bohn}
\affiliation{Escola de Ci\^encias e Tecnologia, Universidade Federal do Rio Grande do Norte, 59072-970, Natal, RN, Brazil } 
\affiliation{Centro Brasileiro de Pesquisas F\'{\i}sicas, Rua Dr. Xavier Sigaud 150, Urca, Rio de Janeiro, RJ, Brazil}
\author{Rubem Luis Sommer} 
\affiliation{Centro Brasileiro de Pesquisas F\'{\i}sicas, Rua Dr. Xavier Sigaud 150, Urca, Rio de Janeiro, RJ, Brazil}
\author{Gianfranco Durin}
\affiliation{INRIM, Strada delle Cacce 91, 10135 Torino, Italy}
\affiliation{ISI Foundation, Viale S. Severo 65, 10133 Torino, Italy}
\author{Stefano Zapperi}
\affiliation{CNR-IENI, Via R. Cozzi 53, 20125 Milano, Italy}
\affiliation{ISI Foundation, Viale S. Severo 65, 10133 Torino, Italy}
\author{James P. Sethna}
\affiliation{LASSP, Physics Department, Clark Hall, Cornell University, Ithaca, NY 14853-2501}
\date{\today}

\maketitle

{\bf 
We report the measurement of {\it multivariable}
 scaling functions for the temporal average shape
of Barkhausen noise avalanches, and 
show that they are consistent with the predictions of simple 
mean-field theories.
We bypass the confounding factors of time-retarded
interactions (eddy currents) by measuring thin permalloy films, and
bypass thresholding effects and amplifier distortions by applying
Wiener deconvolution. We find experimental shapes that are approximately
symmetric, and track the evolution of the scaling function. We 
solve a mean-field theory for the magnetization dynamics and  calculate 
the form of the scaling function in the presence of a demagnetizing field
and a finite field ramp-rate, yielding quantitative agreement with the experiment.
}

The study of critical phenomena and universal power laws has been
one of the central advances in statistical mechanics of the second half
of the last century, in explaining traditional thermodynamic
critical points~\cite{wilson75}, avalanche behavior near depinning
transitions~\cite{fisher98,doussal09}, and a wide variety of
other phenomena~\cite{sethna01}.
Scaling, universality, and the renormalization group claim to predict all behavior at long length and time scales asymptotically close to critical points. In most cases, the comparison between
theory and experiments has been limited to the evaluation of the critical exponents of the power law distributions
predicted at criticality. An excellent playground of scaling
phenomena is provided by systems exhibiting crackling noise, such as the Barkhausen effect 
in ferromagnetic materials \cite{durin06}. Here we focus on the average functional form
of the noise emitted by avalanches --- the  temporal average avalanche shape \cite{sethna01}.

This avalanche shape has been measured for earthquakes~\cite{mehta02}
and for dislocation avalanches in plastically deformed metals~\cite{chrzan94,laurson06},
but the primary experimental and theoretical focus has always been
Barkhausen avalanches
in magnetic systems~\cite{kuntz00,durin02,mehta02,durin06,colaiori08}. 
Theory and experiment agreed well for avalanche sizes and durations, but the 
strikingly asymmetric shapes found experimentally in ribbons~\cite{durin02},
disagreed sharply with the theoretical predictions, for which the asymmetry in the scaling shapes under time reversal was at most very small~\cite{mehta02,sethna01}. (We note that the relevant models are not microscopically time-reversal invariant; temporal symmetry is thus emergent). 
Doubts about
universality~\cite{sethna01} were resolved when eddy currents
were shown to be responsible for the asymmetry, at least on short
time scales~\cite{zapperi05}, but the exact form of the asymptotic universal scaling
function of the Barkhausen avalanche shape still remained elusive.

In this paper, we report an experimental study of Barkhausen noise in permalloy thin films, where a careful study of the average avalanche shapes leads to symmetric shapes, undistorted by eddy currents (which are suppressed by the sample geometry). We provide a quantitative explanation of the experimental results by solving exactly the mean-field theories for two general models of magnetic reversal: a domain wall dynamics model \cite{alessandro90} and the random-field Ising model \cite{sethna93}. The two mean-field theories are
shown to be equivalent, and allow us to compute the average avalanche temporal shape as a function of typical experimental control parameters such as the field rate and the demagnetizing factor.

The relevant statistical information encoded in the Barkhausen noise could in principle be derived from 
the joint two-point time-velocity distribution $G_{c,k}(V,t; V',t+\Delta)$, yielding the conditional probability
that the noise at time $t+\Delta$ is equal to $V'$ if it was equal to $V$ at time $t$. Here $c$ is the
external field rate and $k$ the demagnetizing factor. The standard avalanche distributions can be derived from this parent distribution, e.g. the duration distribution is given by 
\begin{equation}
P(T)=-\int_{0}^{\infty}dV\partial_\Delta G_{c,k}(0^+,t; V,t+\Delta)|_{\Delta=T}.
\end{equation}
The renormalization group makes use of an emergent scale invariance for Barkhausen noise. Here,
the two-point time-velocity distribution, 
when coarse-grained in time by a factor $b$ and rescaled downward in velocity
by a factor $b^x$, rescales at long durations to itself: 
\begin{equation}
G_{c,k}(V,t; V',t+\Delta)=b^{-2x}G_{\bar{c},\bar{k}}(V/b^x,t/b \,; \, V'/b^x,(t+\Delta)/b),
\end{equation}
where $\bar{c}=b^y c$ is the rescaled field rate, $\bar{k}=b^w k$ the rescaled demagnetizing factor and $x,y,w$ are universal scaling exponents. 
Repeating $n$ rescalings until $\Delta/b^n=1$ leads to a universal scaling form
\begin{equation}
G_{c,k}(V,t ;  V', t+\Delta) = \Delta^{-2x}{\cal G}(\Delta^{-x}V,\Delta^{-x}V',\Delta^{y}c/c_0,\Delta^{w}k/k_0).
\end{equation} 
where ${\cal G}$ is a universal {\it multivariable} scaling function and $c_0$ and $k_0$ are the small scale
values of the field rate and demagnetizing factor, respectively. 

Universal scale invariant forms can then be derived for all statistical quantities of interest including the temporal average shape.  For avalanches of duration $T$, the average shape is defined as the average velocity for avalanches that begin and end at $V=0$ in a duration $T$. It has an associated universal scaling form,
\bea
\langle V_{c,k}(\Delta|T)\rangle &=&\int dV' V' G_{c,k}(0^+,0; V',\Delta)G_{c.k}(V',\Delta; 0,T)/\int dV' G_{c,k}(0^+,0; V',\Delta)G_{c,k}(V',\Delta; 0,T)\nonumber\\
&=& T^{x} {\cal V}(\Delta/T,(k/k_0)T^{w},c/c_0T^{y}),
\eea
where ${\cal V}(\lambda,K,C)$  is a {\it universal} scaling function, dependent on the rescaled time 
$\lambda\equiv \Delta/T$, the rescaled demagnetizing factor $K\equiv (k/k_0)T^{w}$, 
and the driving field $C \equiv (c/c_0) T^y$. It is universal in the 
sense that in the scaling regime does not depend on microscopic features of the material, 
so long as the system is at a critical point. 

We record the Barkhausen noise by a standard inductive technique on a 1~$\mu$m thick permalloy film, with
polycrystalline structure (see methods section for details on the measurement and on sample preparation).
The Barkhausen noise is composed by a series of intermittent pulses, due to avalanches in the magnetization,
combined with background instrumental noise. In most crackling noise phenomena, avalanches are usually identified by setting a threshold above the background noise  $V_{\mathrm{th}}$. This method works well if the signal to noise
ratio is high, but can induce spurious effects otherwise \cite{laurson09}. Our thin films have a correspondingly
weak signal, making the $V_{\rm th}$ inappropriate. In addition, 
the measurement apparatus has a response function that distorts the original pulses.
We instead extract the pulses using Wiener deconvolution \cite{book-nr}, which 
optimally filters the background noise and bypasses the use of thresholds (see Fig.~\ref{fig:2}).

The universality class of the Barkhausen noise
in a sample is usually identified by measuring the voltage distribution, the power spectrum and
the distributions of avalanche sizes $S$ and durations $T$
\cite{durin00}. The present experiments display characteristic features of the mean-field universality class: 
we observe a field-rate dependence of the voltage distribution and of the avalanche size and
duration distributions that is in excellent agreement with mean-field predictions
(see Fig.~\ref{fig:0}(a), (b) and (d), noting in each case the three variable scaling implied by the
field rate and demagnetizing factor). 
Furthermore, the power spectrum and the conditional average size of an avalanche of 
duration $T$ are rate independent and consistent with
mean-field predictions (see Fig.~\ref{fig:0}(c)). 
Finally, we focus on the measurement of the average avalanche temporal shape, considering all the avalanches
of a given duration $T$ and averaging the voltage signal at each time step $t$. (In practice,
we average the avalanches using duration bins centered at $T$ and with geometrically increasing sizes). Fig.~\ref{fig:1}(a) shows the resulting nearly symmetric temporal average avalanche shape, which starts out
parabolic and then flattens as the duration of the avalanche increases
(cf.~Fig.~\ref{fig:1}(c)). 

It has been argued that dipolar magnetic fields are sufficiently long-ranged that mean-field theory should be quantitatively applicable in three dimensional samples~\cite{zapperi98}. 
Similar considerations apply to many of the systems exhibiting crackling noise, such as dislocation-mediated plasticity~\cite{zaiser06} and  earthquakes~\cite{fisher97} where long-range interactions are provided by elastic strains. Our films are thinner than polycrystalline ribbons known to exhibit mean-field behavior~\cite{durin00}, but thicker
than previously studied 2D films~\cite{kimnat}. The classic mean-field theory for domain wall
depinning is the single-degree of freedom ABBM model~\cite{alessandro90}, 
which treats the domain wall as a rigid object at position $X(t)$,
advancing in a random pinning field statistically chosen as a {\em random walk}
in $X$,
\bea
\Gamma\frac{dX}{dt}=ct-k X +W(X),
\label{eq:abbm}
\eea
with $\langle W(X)\rangle = 0$ and $\langle W(X)W(X')\rangle = |X - X'|$ (Brownian noise) and where $\Gamma$ is the damping coefficient. The ABBM model predicts that the avalanche size and duration
distributions decay as power laws with rate dependent exponents \cite{colaiori08}. The exact form
of the scaling functions for these distributions, including the cutoff to the power law behavior, can be computed in the limit $c=0$ \cite{colaiori08}. To compare with experiments for $c>0$, we resort to numerical integration of Eq.~\ref{eq:abbm}, obtaining close agreement, as displayed in Fig.~\ref{fig:0}.

The average avalanche shape in the ABBM model has not been
extensively explored numerically, but an approximate analytical
calculation~\cite{colaiori04,colaiori08} gave a lobe of a sinusoid
${\cal V}(\lambda) \simeq \sin(\pi \lambda)$.  In that calculation, the approximation $\langle V(X(t),T)\rangle \simeq V(\langle X(t)\rangle, T)\rangle$ was made.  
(That is, the fluctuations in the growth of the avalanche size with time 
were neglected, leading to a slight distortion of the predicted temporal average avalanche 
shape.)
We avoid this approximation by using a transformation due to Bertotti~\cite{bertottibook}  
to a time-dependent noise with a variance proportional to the avalanche velocity $V$
\bea 
\frac{dV}{dt} = c-k V + \sqrt{2V}\xi(t),
\label{eq:abbm2}
\eea
which can be solved in the Stratonovich interpretation
in the limit $c=0$, $k=0$.
After defining a new variable $Y=V^{1/2}$, we have a stochastic equation that we may solve explicitly. 
Utilizing the resulting probability for the first-return to the origin, we calculate
the temporal shape~\cite{baldassarri03,colaiori08}~(cf.~Supplementary Information) :
\bea
\langle V_{c=0,k=0}(t|T)\rangle = \frac{1}{8} t \left(1-\frac{t}{T}\right),
\eea
where in the future we will use $t$ instead of $\Delta$ to denote the 
time elapsed from the beginning
of the avalanche. We have numerically verified that the average shape is indeed an inverted parabola (and not the lobe of a sine wave \cite{colaiori08}).

Interestingly, an inverted parabola was reported \cite{kuntz00,mehta02} in numerical studies of the (nucleated) random-field Ising model in the mean-field limit (the `shell' model) \cite{sethna93}. Front depinning and nucleated transitions have
different upper critical dimension and different short-range critical exponents, and it was conceivable that their
mean-field theories had different average shapes, albeit sharing critical exponents. 

Upon closer examination, these two models are {\it the same} in the continuum limit.
The shell model has a set of interacting spins $M_i=\pm 1$ with random fields $h_i$
distributed by a Gaussian, interacting with a strength $J/N$ with all other
spins; each spin flips when the net field it feels, $h_i + H(t) + J /N \sum_j S_j$
is positive. Here, $H(t) = H_0 + c t$ is the external field, increasing with rate $c$,
and the spin feels the magnetization $M = (1/N)\sum_j S_j$ both through the
infinite-range coupling $J/N$ and the demagnetizing factor $k$. A shell-model avalanche
proceeds in parallel, with a shell of $V_n$ unstable spins flipping at the $n-$th time-step,
then triggering a new set of spins $V_{n+1}$ to flip:
\begin{equation}
\label{eq:ShellModelV}
V_{n+1} = P(2JV_{n}),
\end{equation}
where $P$ is the Poisson distribution for the set of spins $V_{n+1}$ to be included in the
range $\{f,f+2JV_{n}\}$. For large $V$, we may approximate the Poisson distribution
as a Gaussian $P(V) = V+\sqrt{V}\xi $, leading to $V_{n+1}-V_n\simeq - \tilde k V_n+ \sqrt{J V_n}\xi(t)$
which is clearly a discretized version of Eq.~\ref{eq:abbm2}, where $\tilde k =1-2J$ is the distance to the critical threshold.  This equivalence, in retrospect, provides an alternative explanation for the origin of Brownian noise statistics for the ABBM domain wall potential \cite{zapperi98}.

Finally, we consider the effects of the two main physical perturbations, 
the driving field $c$ and the demagnetizing factor $k$. The transformed Eq.~\ref{eq:abbm2} now takes the form
\begin{equation}
\frac{dY(t)}{dt}=\frac{c}{Y}-kY(t)+\frac{1}{\sqrt{2}}\xi(t).
\end{equation}
Under rescaling $\sqrt{V}= Y \rightarrow b^{-x/2} Y$, $t\rightarrow t /b$ and for uncorrelated Gaussian white noise we have $\xi\rightarrow b^{1/2}\xi$. 
By balancing the noise with the left-hand side, we find $x=1$; thus the driving field is marginal ($y=0$), 
and the demagnetizing factor is relevant with scaling dimension $w=1$.
The demagnetizing factor $k$ sets the
characteristic maximum of the avalanche size and duration. The two-point probability function in the case $c=0$ can be found exactly~\cite{vankampenbook} and the functional form of the resulting shape is,
\bea
\label{eq:CorrectionToScaling}
\langle V_{c=0,k}(t|T)\rangle &=& \frac{1}{2k}\frac{(e^{2k (T-t)}-1)(e^{2k t}-1)}{e^{2k T}-1},
\eea
and hence the scaling function is
\begin{equation}
{\cal V}(\lambda, K, 0)=\frac{1}{K}\frac{(e^{K(1-\lambda)}-1)(e^{K\lambda}-1)}{e^{K}-1}.
\end{equation}
Intuitively, the form of the scaling function leads to a {\it flattening} of the average shape at long times because
$k$ acts to cap the velocities. 
As it can be seen in Fig.~\ref{fig:1}, this behavior is verified by our experiments.

The effect of the driving field $c$, a marginal field at mean-field, can be calculated exactly at $k=0$ using a result from Ref.~\cite{bray00}: for an absorbing boundary condition at $V =0$, the 
two-points probability function for the Brownian motion in a logarithmic potential is $G_{c,k=0}(V,t ; \epsilon, 0) = \frac{4}{\Gamma[1+\nu]}(4t)^{-(1+\nu)}\epsilon^{2\nu}V e^{-V^2/(4t)}(1+O(\epsilon^2))$, where $\nu=(1-c)/2$. Using this expression, the scaling form is reduced in magnitude but remains parabolic, with ${\cal V}(\lambda, 0,C)=(1/8)(1+C)\lambda(1-\lambda)$.

In conclusion, we have used multivariable universal scaling functions to study average avalanche temporal shapes in Barkhausen noise. By analyzing thin films, we have bypassed the eddy current effects that have long frustrated  complete comparison between theory and experiment. We unify the two rival mean field theories, calculate  both the mean-field scaling function for the temporal average shape and the effects of the demagnetizing factor and 
the field rate.  By utilizing optimal Wiener filtering techniques we allow an unbiased capturing of the shapes, allowing us to report almost symmetric shapes, yielding excellent agreement with our theoretical predictions.

\section*{Methods}

\subsection*{Sample preparation and experimental measurements}

A 1~$\mu$m thick ferromagnetic film with nominal composition
Ni$_{81}$Fe$_{19}$ (permalloy) is deposited by magnetron sputtering
from a commercial target, on a glass substrate covered by
a 2~nm thick Ta buffer layer. The deposition is performed
with the substrate moving at constant speed through the plasma in
order to improve the film uniformity, with the following parameters:
base vacuum of $10^{-7}$~Torr, deposition pressure of 5.2~mTorr
(99.99~\% pure Ar at 20~SCCM constant flow), and 65~W RF power. The
deposition rate of 0.28~nm/s is calibrated with x-ray diffraction,
which also confirms the polycrystalline character of the film.
Quasi-static magnetization curves obtained
with a vibrating sample magnetometer indicate isotropic in-plane magnetic properties with an out-of-plane
anisotropy contribution, a behavior related to the stress stored in
the film, and to the columnar microstructure ~\cite{hubertbook,viegas01,santi06}.

Barkhausen noise time series are obtained using the inductive
technique in an open magnetic circuit. The sample has dimensions 10~mm $\times$ 4~mm. Sample and pickup coils are inserted in a long solenoid with compensation for
the effects of the border. The sample is driven by a triangular driving magnetic field with amplitude
high enough to saturate the film ($\pm$ 15~kA/m). The driving field frequency is varied in the range
0.05~Hz-0.4~Hz. Barkhausen noise is detected by a 400~turn sensing coil (3.5~mm long, 4.5~mm wide, 1.25~MHz resonance frequency) wound around the central part of sample. A second pickup coil, with the same cross
section and number of turns, is adapted in order to compensate the
signal induced by the magnetizing field. The Barkhausen signal is
then amplified and filtered with a 100~kHz low-pass filter, and finally acquired data at 0.2 and 4~MSamples/s. The data at different rates are statistically very similar after Wiener filtering (spurious peaks are present in the high frequency regime of the spectrum, independently of the rate), with very small differences at short durations and small voltages. Wiener filtering is practically more efficient at large rates, for uncorrelated noise, but the noise amplitude remains larger than at low rates. Thus, the optimal use of this filter depends on the observables studied: For the distributions in Fig.~\ref{fig:0}, in order to gain more accuracy at short durations and small avalanches sizes by reducing the noise level, the low rate is used; For the average temporal shapes, in Fig.~\ref{fig:1}, being more important to capture variations of the signal at small timescales by a more accurate Wiener filter, the large rate is used. The time series is acquired just around the central part of the hysteresis loop near the coercive field, where the domain wall motion is the main magnetization
mechanism ~\cite{bertottibook, bohn2007}. By using a thin film, we have
removed the confounding effects of eddy currents, whose time scale~\cite{zapperi05} for
the present sample is estimated to be $T_p\sim 0.04~\mu$s, much smaller than the avalanche durations studied.

\subsection*{Data analysis}

\begin{figure}[tbh]
\includegraphics[width=0.96\textwidth]{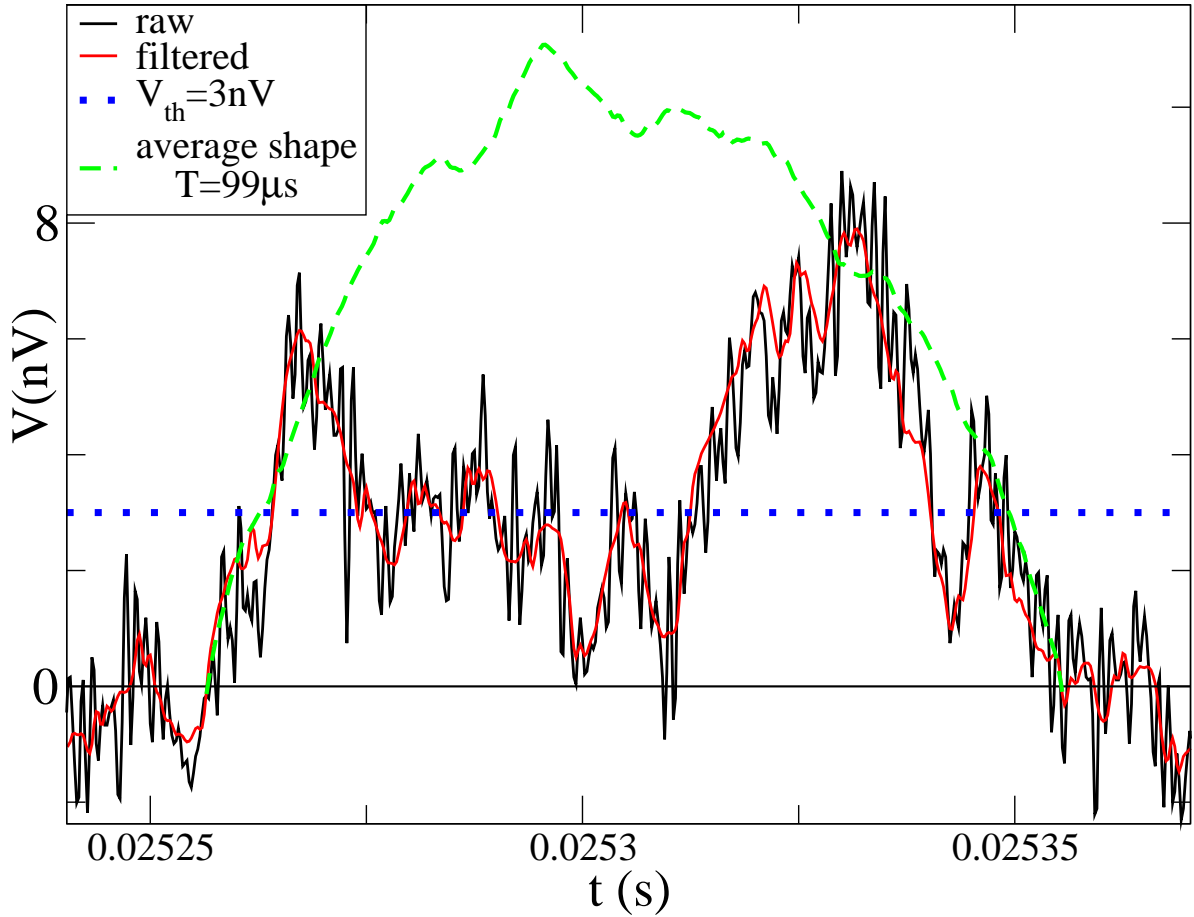}
\caption{{\bf Extracting average shapes from noisy data by Wiener
deconvolution}. Time-series data (jagged line) is traditionally separated into
avalanches using a threshold $V_{\mathrm{th}}$ set above the instrumental
noise (dotted blue line) -- here breaking one avalanche into a few pieces.
We instead do an optimal Wiener deconvolution (smoothed red curve, see text), allowing the use of a zero
threshold (solid black line) which avoids distortions of the average shape and also gives
more decades of size and duration scaling. Averaging over all avalanches
with this duration gives us $\langle V(t,T)\rangle$ (dashed green curve).}
\label{fig:2}
\end{figure}

\begin{figure}[tbh]
\centering
\includegraphics[width=1.\linewidth]{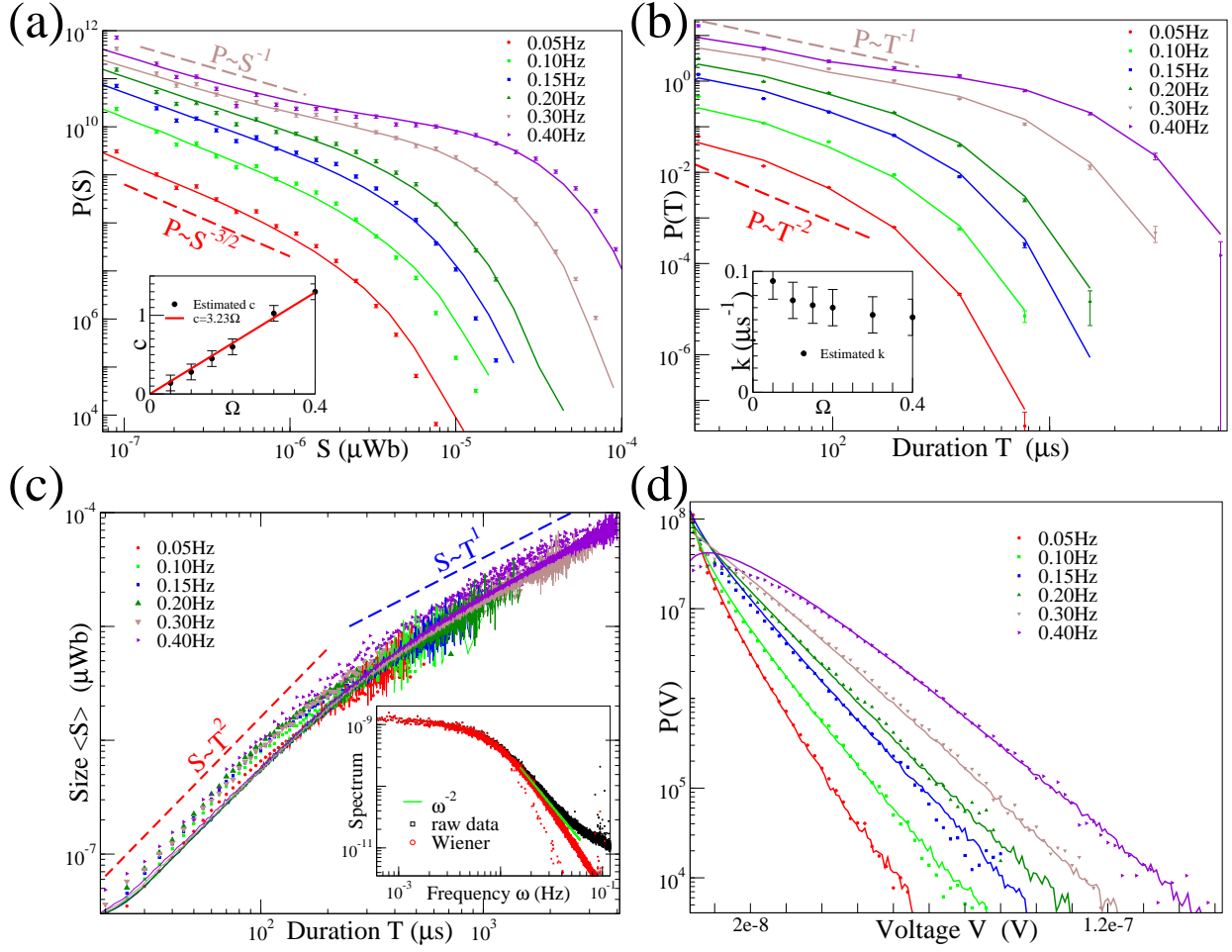}
\caption{\scriptsize{\bf Multivariable scaling; experiment vs.\ mean-field 
theory.} In each main graph, the symbols represent
data taken at various sweep frequencies $\Omega$ for the
external field. The lines represent mean-field predictions extracted
from ABBM simulations (cf. Eq.~\ref{eq:abbm2}) with parameters $c$ and $k$ shown in the insets
of~(a) and~(b), taking similar number of samples with the experiment; the sweep rate $c\propto \Omega$ as predicted, and 
the weak dependence of the demagnetizing factor $k$ can plausibly
be attributed to changing experimental conditions.
The distribution of avalanche sizes and durations shown in~(a) and~(b)
depend on three variables (demagnetizing factor, sweep frequency,
and $S$ or $T$); no analytical form is available, but the agreement
with the mean-field simulations is excellent. 
 Avalanches of longer duration merge when ramped at a finite rate,
leading to unusual rate-dependent critical exponents in mean-field theory;
the smaller avalanches have  $P\sim S^{-\tau}$ with $\tau=3/2 - c/2$,
while the briefer ones have $P\sim T^{-\alpha}$ with $\alpha=2-c$.
(Vertical scales are shifted for clarity; vertical units are arbitrary.)
The limiting values of $c=0$ and $c=1$ (corresponding to $\Omega=0.3$)
are shown. The mean avalanche size versus duration is shown in~(c); 
for avalanches below the demagetization cutoff $\langle S\rangle \sim T^{2}$,
while above the demagnetization cutoff $\langle S\rangle \sim T$ is observed, as expected.
Power-law behavior of avalanche sizes versus durations has been observed
before in experiments on systems with short-range interactions, but not before on
materials believed to belong to the long-range or mean-field universality class~\cite{durin06} (presumably because of eddy currents). The observed systematic vertical shift with rate of the experimental curves is due to the presence of instrumental noise, not included in the simulations. The inset shows the power
spectrum for $\Omega = 0.05$~Hz, which has a behavior consistent with $\omega^{-2}$,
as predicted by mean-field theory. Finally, in~(d) the the voltage probability
distribution is consistent with the mean-field prediction
$P(V|k,c)\propto V^{c-1} \exp{(-k V)}$, another three-variable scaling form. The observed deviations at small voltages are due to the presence of instrumental noise, being approximately gaussian with $\sigma\approx18nV$.}
\label{fig:0}
\end{figure}

\begin{figure}[h!]
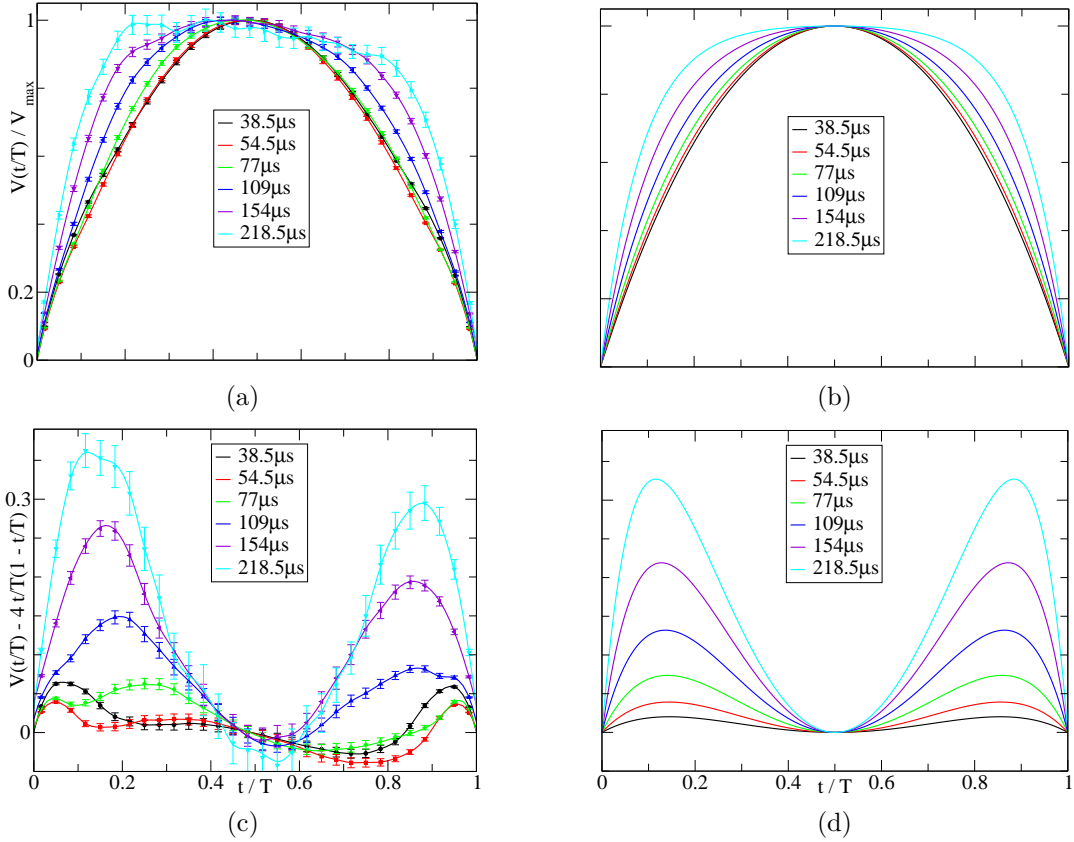

\centering
\subfigure[]{\begin{minipage}[t]{0.47\linewidth}
\includegraphics[width=0.81\linewidth]{fig1_sub1.eps}\vspace{0.2cm}
\end{minipage}}
\subfigure[]{\begin{minipage}[t]{0.47\linewidth}
\includegraphics[width=0.81\linewidth]{fig1_sub2.eps}
\end{minipage}}
\subfigure[]{\begin{minipage}[t]{0.47\linewidth}
\includegraphics[width=0.82\linewidth]{fig1_sub4.eps}
\end{minipage}}
\subfigure[]{\begin{minipage}[t]{0.47\linewidth}
\includegraphics[width=0.82\linewidth]{fig1_sub3.eps}
\end{minipage}}

\caption{{\bf Experiment and theory: Average shapes and 
scaling.} (a)~The temporal average avalanche shape 
for different avalanche durations $T$, rescaled to unit height and duration.  The 218.5~$\mu$s curve, for example, represents the average over $658$ avalanches, with durations between 181~$\mu$s and
256~$\mu$s, binned in $50$ equal time intervals, each of size $4.37~\mu$s .
Notice the symmetric, parabolic shape for short durations, and the flattening
for longer durations. The small residual asymmetry remaining is plausibly
attributable to amplifier-induced distortions that were not possible to correct for. The impulse response functions of the individual instruments (amplifier, low-pass filter and sensing coils) were measured subsequent to taking the data, and used to deconvolve the resulting signal~(cf.~Supplementary Information). (b)~Mean-field calculation for the average shape
including a demagnetizing factor $k$ as in Eq.~\ref{eq:abbm}. The shape is an inverted parabola,
$4 t/T (1-t/T)$ for short durations and flattens at long durations.
Here, $k$ is $0.05~\mu$s$^{-1}$, slightly different
from the value used in Fig.~\ref{fig:0}. This difference is expected since our analytical form is only
available for $c=0$; simulations at $c,k$ given in the insets  of Fig.~\ref{fig:0}a and b respectively, yields good,
albeit noisy, agreement with experiment.
(c,d)~Deviations from the inverted parabola. Note the quantitative
agreement between experiment and theory.}
\label{fig:1}
\end{figure}

To address the low signal from the thin film, we analyze the data by
using Wiener deconvolution \cite{book-nr}.  
The form of the output signal is assumed to be of the form $V_{out}(t)=(h\star (V +n))(t)$ where $h(t)$ denotes the impulse response function, $V(t)$ the original microscopic signal, and $n(t)$ the background noise. Given an estimated impulse response Fourier series $\tilde{h}(f)$, an estimated deconvolved noise power spectrum $|\tilde{n}(f)|^2$ and a theoretically expected frequency spectrum for
the deconvolved signal $|\tilde{v}(f)|^2$, the filtered data $V(t)$ is the inverse 
Fourier transform of
\bea
\label{eq:Wiener}
\tilde{V}(f) = \frac{\tilde{V}_{out}(f)}{\tilde{h}(f)}\frac{|\tilde{v}(f)|^2}{|\tilde{n}(f)|^2+|\tilde{v}(f)|^2}.
\eea
Here, the impulse function is estimated by measuring the impulse response of the individual instruments composing the apparatus. We note that if the signal has a high resolution, a detailed knowledge of the impulse response
function is needed to remove the effect of the instrument.
 As a signal function,
we use a power fit $|\tilde{v}(f)|^2\sim f^{-\gamma}$ (see also Supplementary Information). To estimate the noise
contribution $|\tilde{n}(f)|^2$, we measure the power spectrum of the instrumental
noise, recorded without the material, maximizing over several runs. The Wiener deconvolution, 
shown in Fig.~\ref{fig:2} and in the inset of Fig \ref{fig:0}c, smoothens the signal, and removes spurious high frequency oscillations, due to the amplifier and filters used in the experiments. Importantly, this procedure also allows us to avoid the use of thresholds for defining the temporal extent of the avalanches, improving
our estimates for both the scaling exponents and the average avalanche shapes.

\vspace{2cm}

{\bf Supplementary Information} is linked to the online  version of the paper.

\begin{acknowledgments}
We would like to thank F. Colaiori, K. Daniels and K. Dahmen for enlightening discussions. FB would like to thank M. Carara and L. F. Schelp for the experimental contribution and fruitfull discussions. S.Z. acknowledges financial support from the short-term mobility program of CNR. S.P. and J.P.S. were supported by DOE-BES and R.L.S. and F.B. were supported by CNPq, CAPES, FAPERJ and FAPERGS.
\end{acknowledgments}

\section*{Author Contributions}
F.B., G.D., and R.L.S. were responsible for the experiments. F.B. had
primary responsibility for the sample production and characterization,
the experimental setup, and the experimental data acquisition.
R.L.S. was responsible for the design of the sample production and
characterization methods and supervised the experimental design
and execution. G.D. inspired the project, guided the development
of the experimental design, and both supervised and participated in the
data acquisition. S.P., S.Z., and J.P.S. were responsible for the
implementation of the Wiener
filtering methods. The data analysis, partly based on software by S.Z., was
adapted by all three, and then tested, refined, and executed by S.P.
F.B., G.D., and R.L.S. provided key input into the interpretation and
presentation of the data. S.P. was responsible for the theoretical analysis
of the avalanche shapes and for the simulations. S.P. wrote the original text
of the manuscript; all authors contributed to refining and focusing the text.

\section*{Author information}
The authors declare no competing financial interests. Correspondence and requests for materials should be addressed to S.P. (stefan@ccmr.cornell.edu).

\section*{Figure Legends}

\newpage

\section*{Supplementary Information}
\setcounter{page}{1}
\begin{center}
{\bf Beyond power laws: Universality in the average avalanche shape}\\
Stefanos Papanikolaou, Felipe Bohn, Rubem Luis Sommer,
Gianfranco Durin,\\
Stefano Zapperi, and James P. Sethna
\end{center}

In this supplementary information, we
\begin{enumerate}[(A)]
\item Explain in detail how we implemented the
optimal Wiener filtering to extract the magnetic avalanche sizes, durations,
and temporal shapes from a noisy Barkhausen signal.
We show that our methods not only allow for extraction of avalanche shapes,
but provide more faithful extraction of avalanche sizes, durations, and 
velocities from the signal, by comparing filtered and unfiltered analyses
both of experiments and of simulations with added noise.
\item Further characterize our films, and 
discuss issues involved in the 3D-2D crossover, expected in the 
study of thinner films.
\item Explain the theoretical mapping between the two discussed mean-field theories 
and describe the approximation made in the previous average shape 
calculation~(which gave a sine scaling function rather than an inverted 
parabola). In order to demonstrate the technique, we show in detail how to calculate the average shape for
$k=c=0$.
\end{enumerate}

\subsection[(A)]{Optimal Wiener filtering for extracting Barkhausen avalanches}
In this publication we focus on thin magnetic films to avoid the complexities
of eddy currents found in ribbons. Because of the reduced thickness, the induced signal is significantly smaller, substantially decreasing the signal-to-noise ratio. The larger number of turns of pick-up coil compensates for the reduced voltage, but also introduces spurious background effects, mainly in the high frequency regime of the spectrum.
Such effects clearly demand a more sophisticated extraction of the signal from the noise. We found that the most effective results are obtained by applying an optimal Wiener filter, using independent measurements of the spectral properties of the background noise.

 \begin{figure}[tbh]
 \includegraphics[width=0.4\textwidth]{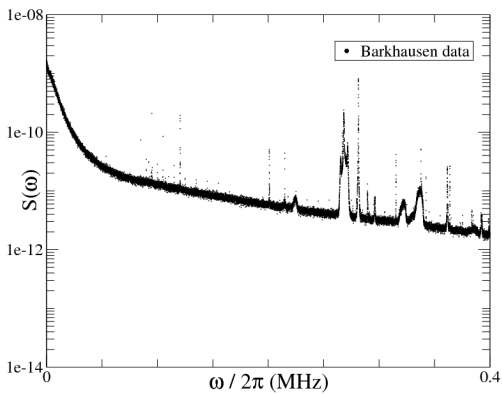}
 \caption{{\bf The average power spectrum of the signal}. The power spectrum at high frequencies, averaged over 200 runs, is dominated by background noise effects which alter the scaling behavior at intermediate sizes and durations.}
 \label{fig:average-signal}
 \end{figure}

 \begin{figure}[tbh]
 \subfigure[]{\includegraphics[width=0.37\textwidth]{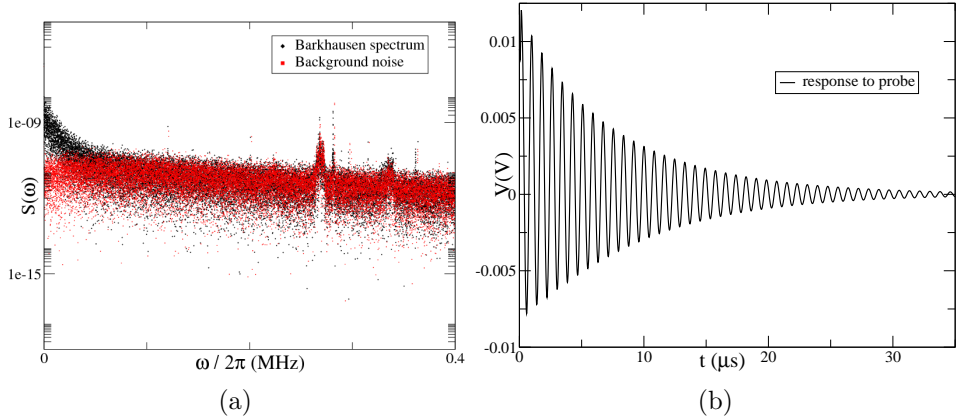}}
 \subfigure[]{\includegraphics[width=0.39\textwidth]{impulse.eps}}
 \caption{{\bf The power spectrum of the instrumental noise after one sweep and the impulse response}. The black points in~(a) are the power spectrum
 of one Barkhausen run, out of 200 total used in this paper, deconvolved with
 the function shown in~(b), based on the measured impulse response
 function for the instrument. The red points in~(a) denote the background deconvolved noise spectrum, which (after deconvolution) resembles white noise at low frequencies. The instrumental response in~(b) is composed of a low-pass filter response, which is a pure exponential decay and a sensing coil response, which amounts to an oscillatory exponential decay. \label{fig:S2}}
 \end{figure}

\begin{figure}[tbh]
\includegraphics[width=0.4\textwidth]{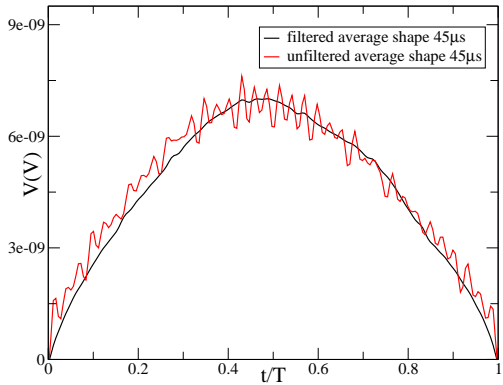}
\caption{{\bf The effect of instrumental noise on the average shapes}. The effect of the background noise oscillations on the form of the average shapes is easily observed in the individual avalanches (Fig.~1 in the main text), and
dramatic in the average avalanche shapes for fixed (non-binned) durations 
(above). After filtering, the oscillations are reduced significantly.}
\label{fig:temp}
\end{figure}

\begin{figure}[tbh]
\subfigure[]{\includegraphics[width=0.4\textwidth]{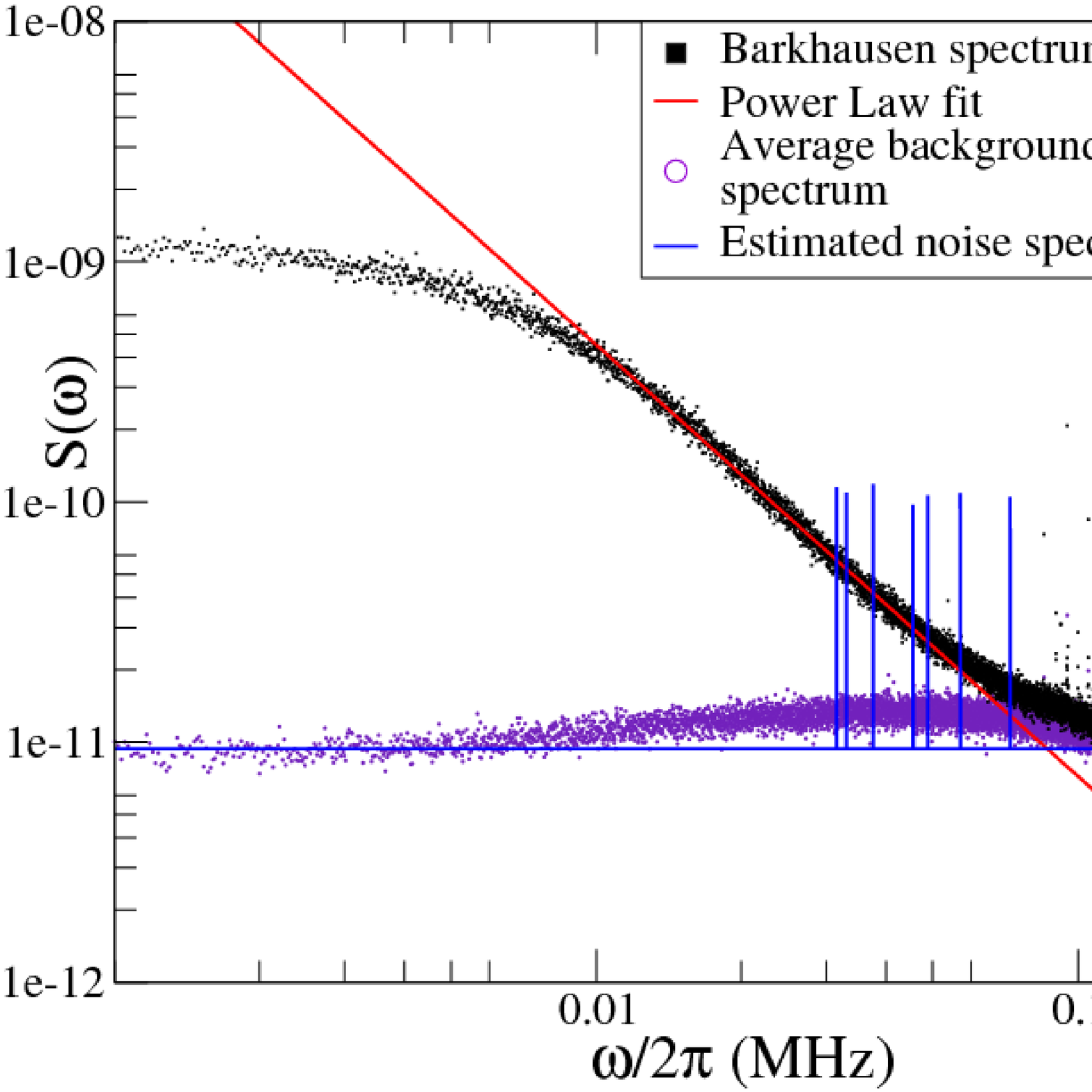}}
\subfigure[]{\includegraphics[width=0.4\textwidth]{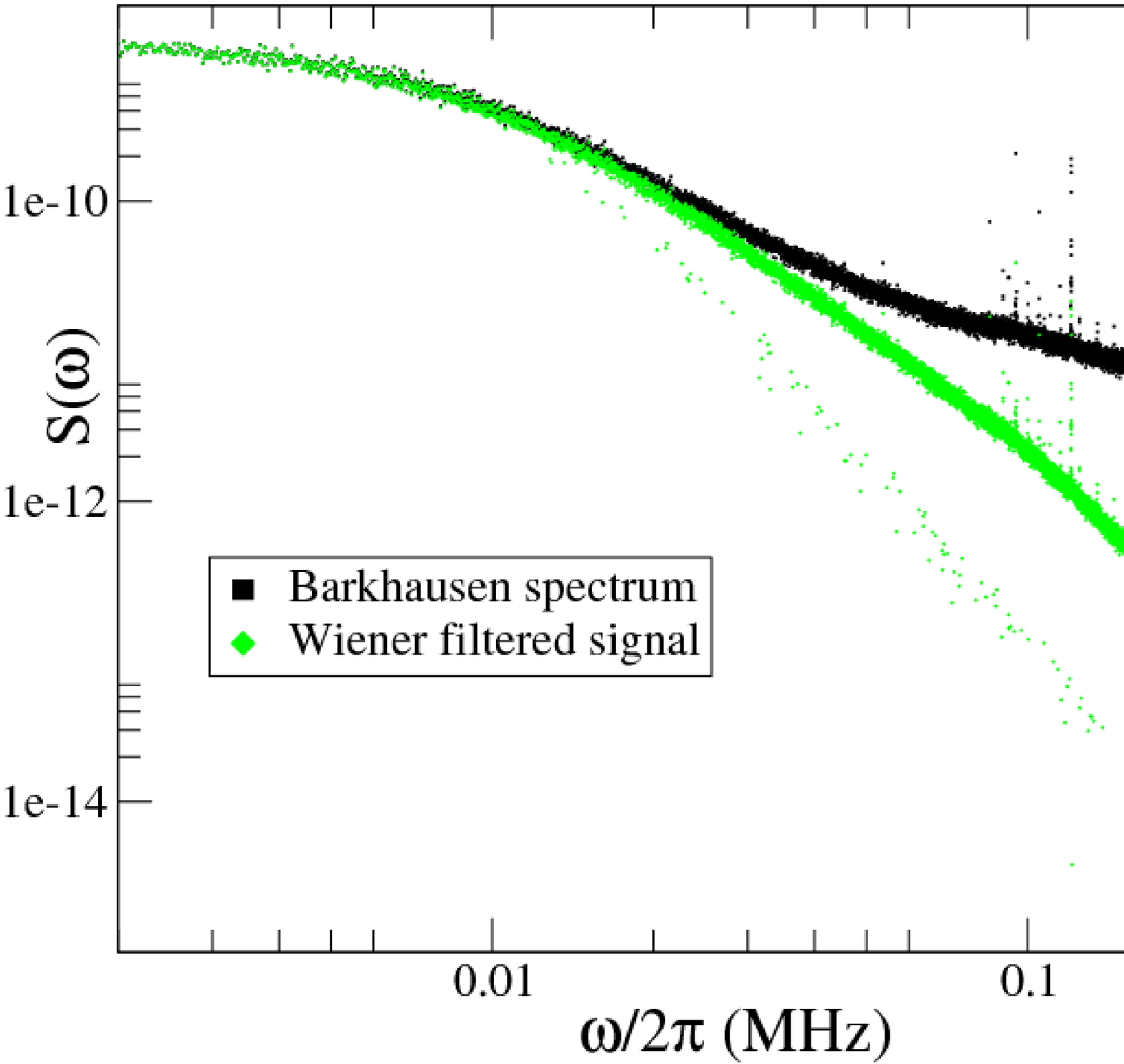}}
\caption{{\bf The power spectrum and Wiener deconvolution.} (a) The black squares denote the power spectrum
of 200 Barkhausen runs for the rate 0.05Hz, deconvolved with the impulse response function of Fig.~\ref{fig:S2}(b). The red curve is a power law fit to the scaling regime of the averaged signal, used in the Wiener deconvolution method. The purple circles denote the average of the background noise spectrum, deconvolved in the same way as the signal, over 200 independent measurements. Finally, the blue curve denotes the estimated noise spectrum, used in the Wiener deconvolution method, which is composed of a white-noise component located at the average amplitude of the background spectrum in the low-frequency regime, and the spurious isolated peaks at high frequencies. Since the positions of the spurious peaks move from run to run, we include them in the noise-estimate whenever the background-noise spectrum of a {\it single} run has larger than $3\sigma_n$, where $\sigma_n$ is the variance of the noise. The form of the signal is plagued with high frequency features that come from artificial background noise elements. (b)  After the Wiener deconvolution, the resulting filtered signal's spectrum averaged over all runs (green diamonds) reduces significantly the distorted features, recovering a large part of the scaling behavior, which is consistent with the mean-field prediction~($S(\omega)\sim \omega^{-2}$). \label{fig:S4}}
\end{figure}

Optimal Wiener filtering is performed in Fourier space, using the
available information about the spectral properties of the signal and
the background noise. The signal is predominant in the low-frequency
portion of the spectrum where, in general, it shows a power law behavior
at intermediate frequencies, going to a constant at low frequencies.
Since we are interested in filtering the high-frequency regime of the
spectrum, we are very careful about estimating the expected signal
$S(\omega)$ at the intermediate frequencies, where scaling takes place.
For this purpose, we use a power law fitting function
$S(\omega)\sim\omega^{-\gamma}$ for our expected
signal~(cf.~Fig.~\ref{fig:S4}(a)). In this way, the very low
frequency regime remains practically unfiltered (since the
spectrum intensity of the signal is much larger than that of the noise
(cf.~Fig.~\ref{fig:S2}(a),~\ref{fig:S4}(a)). The
background noise, after deconvolving the instrumental response
(cf.~Fig.~\ref{fig:S2}(a),~\ref{fig:S2}(b)), is white at
low
frequencies~(cf.~Fig.~\ref{fig:S2}(a) and~\ref{fig:S4}(a)).
The high-frequency portion of the signal and noise are affected by nonlinear
amplifier artifacts~(cf.~Fig.~\ref{fig:S2}(b)), which produce a series of resonance peaks~(cf.~Fig.~\ref{fig:average-signal},~\ref{fig:S2}(a)). 
Their effects on the time signal are spurious oscillations which persist also in the average avalanche shape, as seen in Fig.~\ref{fig:temp}.
We use this information to set our estimate for the noise spectrum
as a sum of two components: i)~a white-noise component, being the mean
of the Gaussian noise ($\sim 10^{-11}V^2Hz^{-1}$), shown in
Fig.~\ref{fig:S4}(a).  ii)~Spurious peaks in the spectrum, with
amplitudes larger than three times the white noise's variance, that are
observed to contaminate the signal at high-frequencies.
Since these noise peaks are quite sharp and slightly shift from run to
run, we use the maximum of these $3\sigma$ events over our 200 independent 
magnetic field sweeps as our estimated noise spectrum (blue curve in
Fig.~\ref{fig:S4}(a)).

With this information on the instrumental noise and on the signal, we
apply Eq.~12~of the main text to the FFT of the time
signal~(cf.~\ref{fig:S4}(a)), and perform the inverse FFT to
get the filtered time signal~(cf.~\ref{fig:S4}(b)). The average
avalanche shapes now appear smooth (see Fig.~2 in the main text).

\begin{figure}[tbh]
\includegraphics[width=0.85\textwidth]{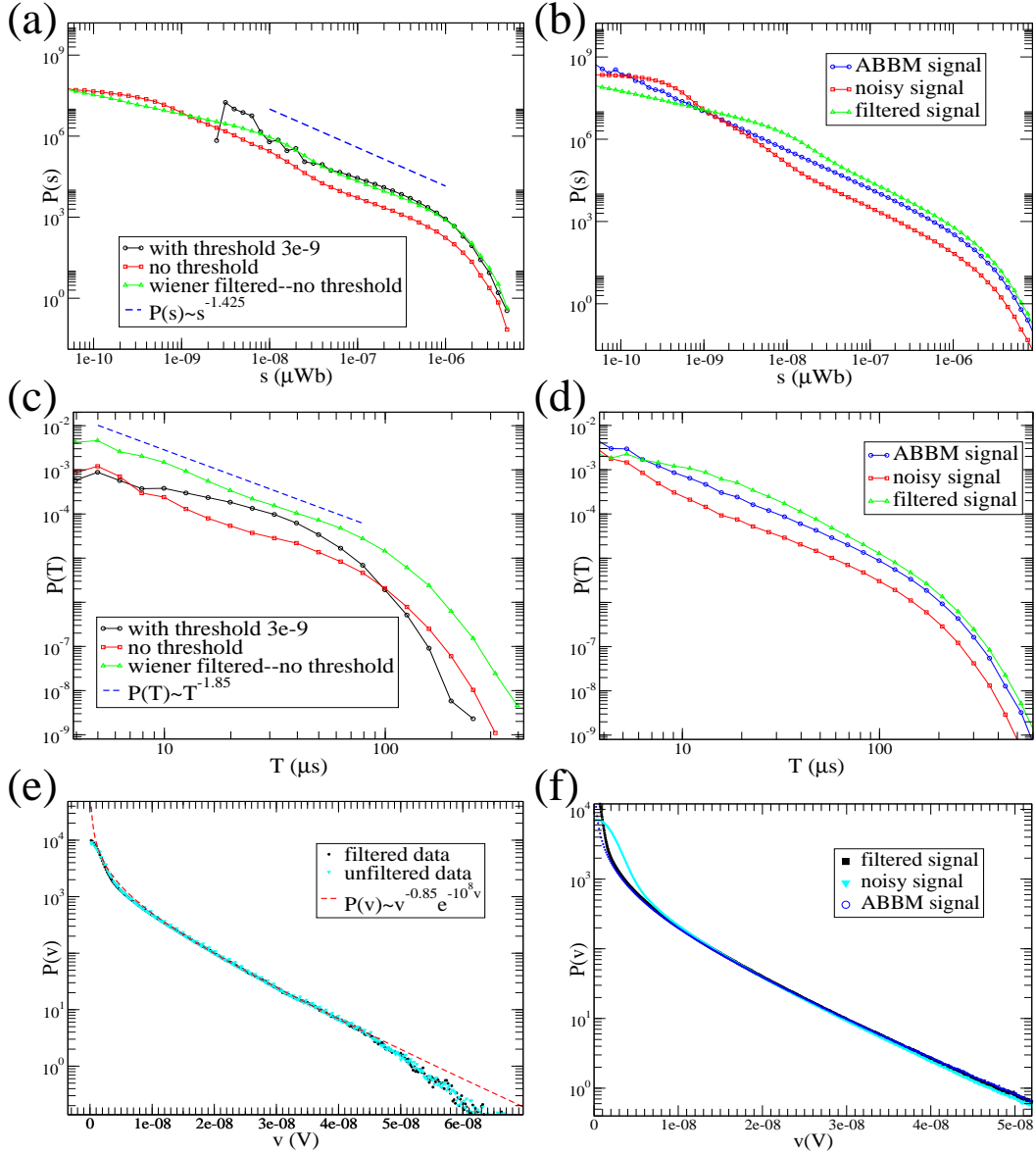}
\caption{
\setstretch{1}
\scriptsize
{\bf Optimal Wiener filtering effects}, tested on experimental
data and simulations with added noise. For each pair, the left panel
is experiment and the right panel is simulated ABBM mean-field data with
$k=0.08$, $c=0.15$, and added noise roughly matching that of the experiment~(at  rate 0.05~Hz),
after deconvolution. (a) and~(b) show the avalanche size distributions;
(c) and~(d) show avalanche duration distributions, and~(e) and~(f) show
voltage distributions. For the experimental size and duration distributions
(a) and~(c), the optimal Wiener filtering (green) appears to allow the 
extraction of almost one decade more than either the raw data (red) or 
data analyzed with the noise threshold depicted in Fig.~1 of the main text.
The theoretical size and duration distributions~(b) and~(d) show the
true ABBM distributions (black), the distributions analyzed with added noise
(red), and the optimally filtered simulations (green). Here we see that the 
filtering allows us to recover many large avalanches that split apart
under the noise, but that the distribution of small and short avalanches
is notably distorted. The voltage distributions~(e) and~(f) are distorted
by the noise and filtering only at low voltages.
\label{fig:PowerLawCombo}
}
\end{figure}

We test our optimal Wiener filtering by studying its effects on 
avalanche sizes, durations, and voltages in Fig.~\ref{fig:PowerLawCombo}. 
We find that the method appears to have significant advantages over
both analyzing the unfiltered data and the alternative method of thresholding 
for extracting smaller avalanches from data (Fig.~\ref{fig:PowerLawCombo}(a) and (c))). 
We also compare to simulated data (cf.~Eq.~6 of the main text) -- we add
Gaussian noise to match that of the experiment after deconvolution, and
then we convolve with the experimentally measured impulse function,
to test for potential systematic errors due to the Wiener filtering. We
find that the smaller and shorter avalanches and the low voltages
(below the noise) are indeed distorted by the process, but that the
recovery of large avalanches is significantly enhanced
by the Wiener filtering process.

\begin{figure}[tbh]
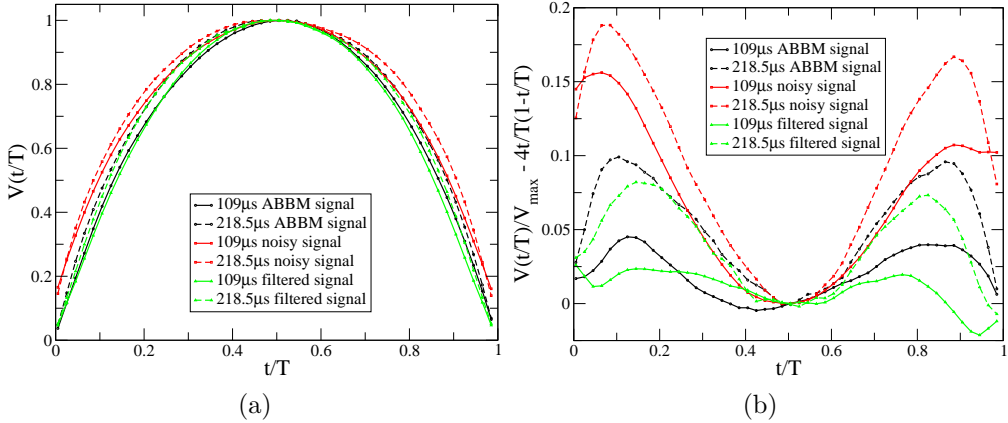

\subfigure[]{\includegraphics[width=0.4\textwidth]{shapes_comp_theory.eps}}
\subfigure[]{\includegraphics[width=0.4\textwidth]{shapescorr_comp_theory.eps}}
\caption{{\bf Simulated avalanche shapes after Wiener deconvolution for the ABBM model before and after filtering}. Two durations of $109~\mu$s and $218~\mu$s shown (where each simulation time-step is taken to be $0.25~\mu$s as in the experiments), before and after filtering the ABBM time series with added Gaussian noise and convolved with the experimentally observed filters. (a) The differences between the ABBM and noisy shapes is significant, but after filtering the behavior is much improved. (b) After adding the noise and convolving with the impulse response function of Fig.~\ref{fig:S2}(b), the average shape
develops a significant asymmetry that resembles the one seen in the experiments. After filtering, this asymmetry disappears completely, since in this case we know the form of the impulse response exactly. In the case of the actual experimental shapes~(cf.~Fig.~2 of the main text), even though filtering reduces the present small asymmetry, it does not disappear completely, presumably because the true impulse function involves non-linear effects that are hard to identify and correct for. Also, note that the noisy signal includes a systematic correction near the edges of the shape~($t\rightarrow0$ and $t\rightarrow T$), due to the presence of background noise.}
\label{fig:shapesthor}
\end{figure}

We also use simulated data to explore possible systematic effects on the
average avalanche shapes.
After adding the noise and convolving with the impulse
response function of Fig.~\ref{fig:S2}(b), the 
average shape develops a significant asymmetry
that resembles the one seen in the experiments, shown in
Fig.~\ref{fig:shapesthor}. After filtering, this asymmetry disappears
completely, since in this case we know the form of the impulse response
exactly. In the case of the actual experimental shapes~(cf.~Fig.~2 of
the main text), even though filtering reduces the present small
asymmetry, it does not disappear completely, presumably because the true
impulse function involves non-linear effects that are hard to identify
and correct for. Also, note that the noisy signal includes a systematic
correction near the edges of the shape~($t\rightarrow0$ and
$t\rightarrow T$), due to the presence of background noise.

\subsection{Films and dimensionality}
Here we characterize our experimental film in more detail, and explain
why it is not surprising that (as clearly shown in Figs~2 and~3 in the main
text) they exhibit critical behavior consistent
with the mean-field theory describing three-dimensional magnets.

\begin{figure}[tbh]
\includegraphics[width=0.4\textwidth]{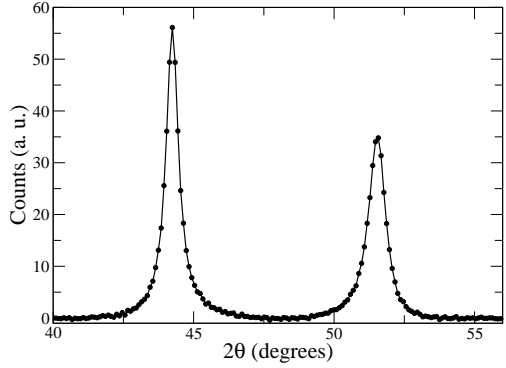}
\caption{{\bf High angle x-ray diffraction pattern confirming the polycrystalline state of the film.} The (111) and (200) permalloy peaks are identified at $2\theta \sim ~ 44.2^{\circ}$, and $2\theta \sim ~ 51.5^{\circ}$, respectively.}
\label{fig:xray}
\end{figure}

\begin{figure}[tbh]
\includegraphics[width=0.4\textwidth]{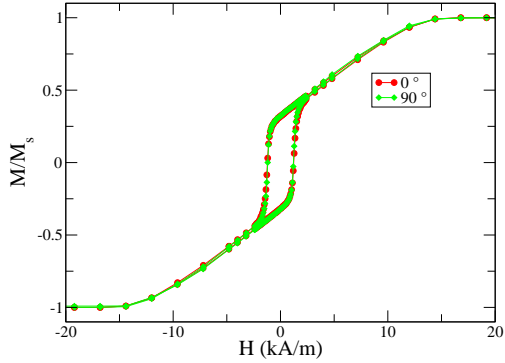}
\caption{{\bf In-plane normalized magnetization curves measured by VSM, obtained along two different perpendicular in-plane directions $\phi = 0^{\circ}$ and $\phi = 90^{\circ}$.} The angle $\phi$ is between the applied magnetic field and direction defined by the motion of the substrate during the deposition. The magnetization curves indicate isotropic in-plane magnetic properties with an out-of-plane anisotropy contribution, a behavior related to the stress stored in the film during the deposition, and to the columnar microstructure \cite{amos08S}.}
\label{fig:mxh}
\end{figure}

Avalanches whose spatial extent is large compared to the thickness of the
film should eventually be described by a two-dimensional theory. Indeed,
for very thin films (below $200$~nm), new universality classes do
appear~\cite{puppin00,kim03,kimnat}. For relatively thick polycrystalline
ribbons (down to $45\mu$m), two of us have previously shown~\cite{durin00}
that Barkhausen noise shows the mean-field behavior 
expected for 3D interface depinning problems with strong dipolar
interactions~\cite{cizeau97} -- the observed avalanches are small enough
compared to the thickness, thus they should not be affected by the asymptotic crossover
to 2D behavior. Our sample has an intermediate thickness of
1$\mu$m, and its polycrystalline state is confirmed by the x-ray structure 
shown in Fig.~\ref{fig:xray}. The magnetization curves of Fig.~\ref{fig:mxh}
show the typical shape exhibited by thick films with a weak out-of-plane
anisotropy, due to the stress stored in the film during the deposition
and to the columnar microstructure \cite{amos08S}. The in-plane magnetic
properties are essentially isotropic, with domains of dense stripes
having widths of the order of the thickness, as confirmed by performing
MFM measurements. Such behavior is known to be absent in
the experiments showing 2D behavior~\cite{puppin00,kim03,kimnat}. 

\subsection{Connections between the ABBM \& Shell models}
Here we derive in more detail the connection between the two mean-field
theories (the ABBM and shell models), and then implement in detail the
calculation of the paraboloic average avalanche shape for the case $k=c=0$.

The equation for the ABBM model (Eq.~2 of the main text) can be transformed to a simpler equation after taking a time derivative~\cite{colaiori08S}:
\bea 
\Gamma \frac{dV}{dX} = \frac{c}{V} - k + w(X).
\label{abbm}
\eea
The noise term has delta-correlations in magnetization 
\bea
\langle w(X)w(X') \rangle_{w} = 2D \delta (X-X').
\label{delta}
\eea
Also, $dX/dt=\ups$ and $\ups \frac{d}{dX}\equiv \frac{d}{dt}$.

Eq.~\ref{abbm} belongs to a class of equations of the form
\begin{equation}
\label{eq:one}
\frac{dV}{dX} = f(V) + \xi(X), 
\end{equation}
with
\begin{equation}
\label{eq:two}
\langle\xi(X) \xi(X')\rangle = \sigma_X^2 \delta(X-X').
\end{equation}
The units of $[\xi] = 1/t$, so the units of $[\sigma_X] = \sqrt{X}/t$.
We interpret Eq.~\ref{eq:one} as a rule for incrementing $V$ and $t$ in
discrete steps of size $\Delta X = \epsilon$:
\begin{equation}
\label{eq:three}
V_{X+\epsilon} = V_X + \epsilon \, f(V) + \Xi(X),
\end{equation}
where now $\Xi(X)$ is an uncorrelated Gaussian variable of standard deviation
$\sigma_\Xi = \sqrt{\epsilon} \sigma_X$~\cite{footnote}. This can be seen explicitly;
$\sigma_\Xi$ represents the RMS width of a random walk with noise $\Xi$ of
duration $\epsilon$:
\begin{align}
\label{eq:ContinuousToDiscrete}
\frac{dy}{dX} &= \xi(X) \cr
\langle (x(X+\epsilon) - x(X))^2\rangle &= 
\int_X^{X+\epsilon} d\mu \int_X^{X+\epsilon} d\mu'
	 \langle \xi(\mu) \xi(\mu') \rangle = \epsilon \sigma_X^2,
\end{align}
using Eq.~\ref{eq:two}.

Now let us consider finding $V(t+\Delta t)$,
where $\Delta t$ is large compared to $\epsilon/V$ (so many $m$-steps
are averaged over) but small enough that $V$ doesn't change significantly
during the interval~\cite{footnote}. We do this by repeating Eq.~\ref{eq:three},
$N = V\,\Delta t/\epsilon$ times. Then
\begin{align}
V(t+\Delta t) &\approx V(t) +  f(V)V\, \Delta t 
	+ \sum_{\tilde X = X(t)}^{X(t)+V\, \Delta t} \Xi(\tilde X) .
\end{align}
The sum has $N$ terms, so is a Gaussian uncorrelated
variable with standard deviation given by 
\begin{equation}
\sqrt{N}\, \sigma_\Xi
	= \sqrt{\frac{V \Delta t}{\epsilon}}  \,
		\sigma_X\sqrt{\epsilon}
		    = \sqrt{V \Delta t} \sigma_X.
\end{equation}
If we define a new discrete Gaussian variable $Z$ of standard deviation
$\sigma_Z = \sqrt{\Delta t}\,\sigma_X$, then multiplying $Z$ by $\sqrt{V}$
mimics the sum, so we find
\begin{equation}
V(t+\Delta t) = V(t) + V f(V)\, \Delta t + Z(t) \sqrt{V}.
\end{equation}
If we now change to continuous time, then (analogous to
Eq.~\ref{eq:ContinuousToDiscrete}) $\sigma_Z = \sigma_\zeta \sqrt{\Delta t}$,
so
\begin{equation}
\label{eq:seven}
\frac{dV}{dt} = V f(V) + \sqrt{V} \, \zeta(t),
\end{equation}
yielding Bertotti's result quoted in the main text. Here, $\zeta(t)$ satisfies
\begin{equation}
\label{eq:eight}
\langle\zeta(t) \zeta(t')\rangle = \frac{\sigma_Z^2}{\Delta t} \delta(t-t')
				 = \sigma_X^2 \delta(t-t').
\end{equation}
From Eq.~\ref{eq:seven}, the units of $[\zeta] = \sqrt{X/t^3}$;
from Eq.~\ref{eq:eight}, the units of 
$[\langle \zeta \zeta \rangle] = X/t^3$, matching the units of $[\sigma_X]^2 [\delta(t)] = (X/t^2) [1/t]$.

A nice physical idea exists behind Eq.~\ref{eq:seven}: Namely, the motion $ dX/dt $ is always forward, so it
always uncovers ``new" random forces as the front progresses. Thus the
noise as a function of time must be delta-function correlated (the new
sites being passed are uncorrelated with sites passed at earlier
times). The key question then becomes what the strength of the delta
function is: how does the noise felt by a particle moving fast over a
random environment differ from that felt by a particle moving slowly?
Basically, one is summing over a number of random variables
proportional to $V$, so the noise is increased by a factor of square root
of this number, $\sqrt{V}$.

All the above manipulations have been performed using the convenient Stratonovich 
interpretation. In those cases where we have calculated average shapes using the more physical Ito calculus, the 
results have remained unchanged, up to dimensionless constant factors. We conjecture that the two methods, although different in detail, do not differ in their predictions for universal quantities. 

\subsubsection*{The temporal average avalanche shape as a first return of a random walk to the origin}

Here, we show explicitly, in the simplest possible example, how one can use the concept of the first return of a random walk to the origin in order to calculate the form of the average shape's scaling function. Let's consider the (no-bias, constant-field) limit where $c\rw 0$ and $k\rw 0$, where it becomes equivalent to the so-called shell model:
\bea
\frac{d\ups}{dt} = \sqrt\ups\xi(t),
\eea 
as described in the main text. Changing variables to $x=\sqrt{V}$, this equation becomes an ordinary random walk. We solve for an initial condition at $\epsilon$ with absorbing boundary conditions at the origin, which yields for  $\epsilon\rightarrow0$,
\bea
G_{RW}(x,t;\epsilon,0)=\sqrt{\frac{2}{\pi}}t^{-3/2}\epsilon~x e^{-x^2/(2t)},
\eea
after using the method of images. In the original $V$ coordinates it becomes:
\bea 
G_{S}(\ups,t;\epsilon,0)=\frac{G_{RW}(\sqrt\ups,t;\epsilon,0)}{\sqrt\ups \sqrt\epsilon}.
\eea
The average avalanche shape is defined as the average of the signal $V(t,T)$ at time $t$ for an avalanche of duration $T$. Given the probability distribution $G_S$, the probability of a signal $V(t,T)$ at time $t$ is 
$G_S(V,t; \epsilon,0)G_S(\epsilon,T;V,t) 
= G_S(V,t;\epsilon,0)G_S(V,t;\epsilon,T)
= G_S(V,t;\epsilon,0)G_S(V,T-t;\epsilon,0)$,
using time-reversal and time-translation invariance of our random walk,
where $G_{S}$ has an absorbing boundary at $V=\epsilon$. In other words, the average of $V(t,T)$ naturally is 
\bea
\langle\ups(t,T)\rangle &=& \lim_{\epsilon\rw0}\frac{\int_{0}^\infty d\ups \ups G_{S}(\ups,t;\epsilon,0)G_{S}(\ups,T-t;\epsilon,0)}{\int_{0}^\infty d\ups G_{S}(\ups,t;\epsilon,0)G_{S}(\ups,T-t;\epsilon,0)} \nonumber \\
&\approx& 2T\left(\frac{t}{T}\left(1-\frac{t}{T}\right)\right) + O(\epsilon),
\eea
and therefore, the shape is an inverted parabola.


\begin{thebibliography}{1}
\bibitem{footnote}{Note that $v>0$ during the avalanche, so we don't need
   to worry about $X(t)$ not being monotone; we never re-use the random 
   forces $\Xi(\tilde X)$.}
\end{thebibliography}
\end{document}